\documentclass{elsart}
\usepackage{graphicx}
\usepackage{amssymb}
\begin{document}

\begin{frontmatter}

\title{ Low Energy Analyzing Powers in Pion-Proton Elastic
Scattering}
\author[TUB]{R.~Meier},
\ead{rmeier@pit.physik.uni-tuebingen.de}
\author[TUB]{M.~Cr\"oni}, 
\author[TUB]{R.~Bilger},
\author[PSI]{B.~van~den~Brandt},
\author[TUB]{J.~Breitschopf},
\author[TUB]{H.~Clement},
\author[ARI]{J.~R.~Comfort},
\author[TUB]{H.~Denz},
\author[TUB]{A.~Erhardt},
\author[EDI]{K.~F\"ohl},
\author[JER]{E.~Friedman},
\author[TUB]{J.~Gr\"ater},
\author[PSI]{P.~Hautle},
\author[TRI]{G.~J.~Hofman},
\author[PSI]{J.~A.~Konter},
\author[PSI]{S.~Mango},
\author[TUB]{J.~P\"atzold},
\author[MIT]{M.~M.~Pavan},
\author[TUB]{G.~J.~Wagner},
\author[TUB]{F.~von~Wrochem}
\address[TUB]{Physikalisches Institut, Universit\"at T\"ubingen,
              72076~T\"ubingen, Germany}
\address[PSI]{Paul Scherrer Institut, 5232 Villigen PSI, Switzerland}
\address[ARI]{Arizona State University, Tempe, Arizona 85287-1504, USA}
\address[EDI]{School of Physics, 
              University of Edinburgh, Edinburgh~EH9~3JZ, UK}
\address[JER]{Racah Institute of Physics, The Hebrew University,
              Jerusalem 92904, Israel}
\address[MIT]{Laboratory for Nuclear Science, MIT Cambridge, MA 02139, USA}
\address[TRI]{TRIUMF, Vancouver, British Columbia, Canada V6T-2A3, and
               University~of~Regina, Regina, Saskatchewan, Canada S4S-0A2}

\begin{abstract}
Analyzing powers of pion-proton elastic scattering have been 
measured at PSI with the Low Energy Pion Spectrometer LEPS
as well as a novel polarized scintillator target. 
Angular distributions between 40 and 120 deg (c.m.) were taken
at 45.2, 51.2, 57.2, 68.5, 77.2, and 87.2~MeV incoming pion kinetic
energy for $\pi^+p$ scattering, and 
at 67.3 and 87.2~MeV for $\pi^-p$ scattering. These new measurements
constitute a substantial extension of the polarization data base at 
low energies.  Predictions from phase shift analyses are compared with 
the experimental results, and deviations are observed at low energies.

\end{abstract}
\end{frontmatter}

\section{Motivation}
Important quantities of the strong interaction can be extracted
from $\pi p$ observables: the $\pi$NN
coupling constant, the sigma term of the proton, and the
size of isospin breaking. Currently, there is no agreement on
the value of any of these quantities \cite{chi2000}. 

Recent values of the sigma term, extracted from elastic pion-proton
scattering, are substantially higher than the
classical value \cite{Koch} based on the KH80 phase shift 
analysis \cite{KH80}.  
Compared to extractions of the sigma term from baryon masses 
\cite{Gasser,Borasoy},
some of these new results \cite{Marcello1} imply a strangeness 
contribution to the mass of 
the nucleon of up to 50\%.
This is at variance with
other investigations, for example estimates from the 
nucleon strange meson cloud \cite{Gutsche} or neutrino reactions 
\cite{Bazarko}. 

The size of isospin breaking in the pion-proton system has
been investigated by using three experimentally accessible
reactions: elastic scattering $\pi^{\pm} p \rightarrow \pi^{\pm} p$
and single charge exchange $\pi^- p \rightarrow \pi^0 n$.
If isospin is conserved, the amplitudes for these reactions
are connected by a triangle relation. The violation of this
relation for s-waves at low energies  (energies below $T_{\pi}$~= 
100~MeV) has been tested 
by several groups, leading to contradictory results. While empirical
analyses by Gibbs et al. \cite{Gibbs} and Matsinos \cite{Matsinos}
showed large isospin violation of about 7\%, a recent
work by Fettes and Mei\ss ner lead to just 0.7\% \cite{fettes}.

These discrepancies could be at least partly due to deficiencies 
in the $\pi p$ data base. While a consistent and dense data base 
for differential cross sections \cite{Marcello} and analyzing
powers \cite{Sevior,Hofman1,Hofman2} exists today for energies 
across the $\Delta$ resonance, this is not the case
for energies below 100~MeV. Here, information is still 
missing, and there are regions where existing experimental data are
contradictory; in particular, discrepancies in $\pi^+ p$
differential cross-section data have been identified 
\cite{FettesMatsinos}.
New measurements of $\pi p$ observables at low energies 
aim at providing the missing information and removing
the discrepancies. 

Polarization observables are particularly interesting as they
provide information complementary to differential cross-section
data in the sense that they are sensitive to small amplitudes. 
They could also help to resolve contradictions in the cross
section data base. Until recently, only one set of analyzing powers 
at one $\pi^+$ energy was available below 98~MeV \cite{Wieser}, 
largely due to experimental difficulties.
Since then, $\pi^- p$ analyzing powers have been measured down to 
57~MeV~\cite{Patterson} by the CHAOS collaboration at TRIUMF. 

In this article, we now report on measurements of analyzing powers 
in $\pi^+ p$ scattering at several energies between 45.2 and 87.2~MeV, 
as well as a few $\pi^- $p data points at 67.3 and 87.2~MeV. All data 
were acquired with the Low Energy Pion Spectrometer (LEPS) 
and a novel polarized scintillator target at PSI. 

\section{Experiment}

In pion-proton elastic scattering, a spin 0 projectile is scattered
off a spin 1/2 target particle. The reaction is therefore described by a 
spin-flip and a non-spin-flip amplitude, which gives rise
to three polarization observables in addition to the unpolarized differential
cross section. Of the polarization observables, only the analyzing power is
accessible in single-scattering experiments.
The analyzing power $A_y$ describes the modification of the
differential cross section when the target protons are 
polarized perpendicular to the scattering plane. 
With the scattering angle $\Theta$ and the target polarization $P$,
this modification is given by
\begin{equation}
\frac{d\sigma}{d\Omega}(\Theta,P)=\frac{d\sigma}{d\Omega}(\Theta,P\!=\!0)
\left( 1+P\,A_y \right).
\end{equation}
The analyzing power is determined from this linear dependence 
by measuring the differential cross section for different 
values of the target polarization $P$. The highest
sensitivity is reached when large polarization values with opposite
signs are used. Measurements at other polarization
values (in particular at $P=0$) can be used as systematic checks
for the apparatus and the analysis procedure, as the linear dependence 
on $P$ has to be reproduced. For the determination of the analyzing
power from Eq. 1, no absolute normalization of the measured cross sections
is needed. However, the measurements at different polarizations
have to be appropriately normalized to each other.

A major difficulty with measurements of the analyzing power is
caused by the composition of the polarized target. The only
polarized proton targets available for use in a secondary
pion beam are dynamically polarized solid targets. Therefore, the target
material contains nuclei (typically carbon) besides the protons,
and it is surrounded by liquid helium along with copper and iron walls.
Pion reactions on these materials give rise to large backgrounds
for the pion-proton reaction.
At high energies and backward scattering angles, this difficulty
is overcome by measuring the scattered pion and the recoil
proton in coincidence. For low energies and forward angles, this
technique can not be applied as the recoil energy  
is too low to allow the proton to leave the target. This difficulty 
is the main reason for the lack of information on analyzing powers
below a pion bombarding energy of $T_{\pi}$~= 100~MeV (with the notable
exception of the $\pi^+ p$ measurement at 68.3 MeV 
by Wieser et al. \cite{Wieser}). 
It applies in particular to $\pi^+ p$ scattering, where the interesting 
region with the largest expected energy dependences of the analyzing 
power lies
at forward angles. In the measurements reported here, a polarized
active target was the key to the success with background suppression.
This device gave access to the recoil proton energy deposition in
the target through a scintillation light readout.

The experiment was done at the $\pi$E3 pion beam line at
PSI. It employed the Low Energy Pion Spectrometer LEPS \cite{KF}.
LEPS is a compact magnetic spectrometer consisting of a quadrupole
triplet and a split dipole.  The total length of the particle flight path 
was  approximately 5 meters. The trajectory of traversing particles
was sampled in 6 layers (3 horizontal, 3 vertical) of proportional 
chambers between the quadrupole triplet and the dipole,  as well as in a
drift chamber in the focal plane. 
The remaining volume of the spectrometer was kept under vacuum. 
The nominal solid angle
of the spectrometer was 25 msr. The momentum resolution was better than
5$\cdot$10$^{-3}$ for pions in the momentum range used in this experiment.

The target sample, a block of 18*18*5 mm$^3$ scintillating organic polymer 
doped with free radicals  (TEMPO) \cite{Ben}, 
was dynamically polarized in a field of
2.52 T in a vertical $^3$He-$^4$He dilution refrigerator. The
magnetic field was generated by a superconducting split-pair
Helmholtz coil located above and below the target cell. Polarization
was induced in the sample by microwave irradiation. The 
polarization was measured by using NMR methods. 

Data were collected in two operating modes of the polarized target.
In the first mode, the target was continously polarized at 
the full magnetic field. In the second mode, 
the field was reduced to 1.2~T and 
the target temperature reduced to 60~mK (``frozen spin" mode), 
resulting in a typical polarization decay time of 80h.

A plastic light guide with diameter of 12 or 19 mm (for different 
parts of the experiment) 
transported the scintillation light from the sample in the mixing 
chamber to a photomultiplier outside 
the cryostat. The photomultiplier signal was read out by a 1 GHz
flash ADC which allowed offline definition of integration times. 
The gain of the photomultiplier was monitored by an
LED light pulser system.

The data-taking procedure included frequent changes of the polarization
direction, which  were  achieved by changes of the frequency of the
microwave irradiation. This was done in order to verify the long term 
stability of the experimental system. Typically three changes of the 
polarization direction were performed for each data set at each angle 
and energy setting.  Additional data were taken at zero polarization. 

\section{Analysis}

Calculation of the analyzing power required determination of
the incident pion beam energy $T_\pi$, target polarization $P$, scattering
angle $\Theta$, and relative cross sections $\sigma_{rel}$ for different 
target polarizations. The analyzing power for a specific beam energy 
and scattering angle was then extracted from the linear 
$P-\sigma_{\mathrm{rel}}$
dependence (Eq. 1). The extraction of the required parameters, the
subsequent calculation of the analyzing power, and the determination of
systematic errors is described below.

\subsection{Beam energy}

The incident pion energy $T_\pi$ at the target center was determined from the 
pion momentum at the exit of the $\pi$E3 beamline and the pion energy loss
traversing the material from the beam line exit to the center of the target.

The $\pi$E3 channel had been calibrated to an accuracy of $\Delta p/p$ = 0.2\% 
for a previous experiment \cite{Wieser}. This calibration was verified
in the current experiment by measuring the kinematic
shifts of the outgoing pion momenta for $\pi p$ and $\pi$C elastic scattering 
with a thin (1~mm) polyethylene target and the LEPS spectrometer at various
angles and energies. No deviation from the expected beam momentum was seen
within the accuracy of this consistency check.

The energy loss and spread of the pion beam on the path to the target center 
was calculated by a simulation program which took into account the 
geometry, the materials, the field of the polarizing magnet, and
the vertical dispersion of 5 cm per percent of momentum of the beam at
the target position. The resulting
energy spread ranged from $\pm$0.8 to $\pm$1 MeV,
dependent on the incident momentum and the target configuration. 
The mean energy was determined to an accuracy of better than 0.3 MeV 
from the beam calibration.
 
\subsection{Polarization}

NMR absorption signals in the region of the Larmor frequency of the protons
were taken periodically while the scintillator target was dynamically
polarized. For the absolute calibration of the polarization, 
thermal-equilibrium (TE) signals were taken repeatedly at 
a temperature of 2.17 K in a magnetic field of 2.52 T. 
The dynamic (DYN) polarization
was determined as the product of the known TE polarization and the ratio
of the areas of the DYN and the TE signals. The maximum polarizations were
$P_z^+ = +0.522$ and $P_z^- = -0.507$. The dominant systematic error 
of the target polarization comes from the uncertainty in the 
measurement of the temperature under TE conditions 
($\Delta P_z / P_z$ = 0.035).

Part of the data (sets 1 and 2) were taken in frozen spin mode 
in a holding field
of 1.2 T. The polarization decay time was in the range of 60 to 100~h.
In these cases the polarization was measured in the full polarizing
field of 2.52~T before and after the data taking runs. 
The subsequent analysis assumed an exponential decay of the polarization.  
Typically, the relative polarization decay between measurements was 20\%.

\subsection{Scattering angle}

The mean scattering angle was determined to an accuracy of 0.3 degrees from 
the trajectories of the scattered pions, the position of the spectrometer, 
and the distribution of scattering events over the angular acceptance
of the spectrometer.
The trajectories of outgoing pions in the 
magnetic field of the target were calculated 
by the simulation program mentioned in section 3.1. The spectrometer 
position was determined to better than 0.05 degrees. The mean scattering 
angle and the angular
acceptance for particles transported through LEPS were extracted from 
distributions of the accepted angles measured in the LEPS intermediate focus
and projected back to the target position.
The overall angular acceptance due to the incident beam divergence, 
vertex position on the target, straggling, and the acceptance of
the spectrometer was $\pm$2.4 degrees for measurements with a target
magnetic field of 1.2~T (sets 1 and 2), and $\pm$2.8 degrees for
2.52~T (set 3).

\subsection{Relative cross sections}

Relative cross sections were extracted from the LEPS focal plane spectra.
Fig. 1 shows typical focal plane spectra for different conditions
and polarizations. The plotted quantity is $E_{\mathrm{loss}}$, 
which is defined 
as the difference between the kinematically expected energy of the 
outgoing pion in $\pi p$ elastic scattering and the energy measured
in the spectrometer, corrected for the energy loss in traversed materials. 
Spectrum 1 shows all events detected in the spectrometer. The solid line
represents a measurement with positive target polarization, the dotted
line and the shaded area a negative polarization run. The spectra show
a broad bump at negative $E_{\mathrm{loss}}$ which corresponds to 
elastic scattering off carbon in the target material as well as 
scattering off other nuclei, such as copper, in the target cell.
The $\pi p$ elastic scattering peak around $E_{\mathrm{loss}}$ = 0
sits on a large background from other reactions in the target region. 
In this region (and only there), 
a difference between the positive- and negative-polarization
runs is visible. The edge of the spectrometer acceptance for low energy pions
is reflected by the slope above $E_{\mathrm{loss}}$ = 10 MeV.

The relative normalization of the runs at a given energy and angle 
was achieved by a fit of these spectra, 
outside the $\pi p$ region, of all events detected by the LEPS spectrometer
(without requiring any target information). 
This procedure implicitely takes into account 
wire-chamber and data-taking efficiencies. The systematic error in this
procedure was estimated by variation of the $E_{\mathrm{loss}}$ region 
used in the normalization.

Subplots 2 to 4 of Fig. 1 show $E_{\mathrm{loss}}$ spectra when a software 
coincidence of a spectrometer event and a scintillation light signal from 
the active polarized target was required. The minimum size of the required 
signal (called 'cut' in the following) was increased 
in each of these figures.  This target signal 
requirement removed a major fraction of the 
background under the $\pi p$ peak. 

To ensure that the cut was equivalent
for all runs at various polarizations used for the 
calculation of one analyzing power data point, a run-to-run calibration
of the target signal amplitude was required.
The calibration was done by 
using the $\pi$C elastic scattering peak. This peak was independent
of polarization, and so the normalized yield in this peak had to be, 
within statistics, identical for all runs. Therefore the signal size was 
calibrated by requiring the same cut-dependent normalized yield in the $\pi$C
region for all runs. The calibration was expressed as a dependence
of the equivalent cut for one run versus the cut in a reference run. 
This dependence was found to be well 
described by a linear function. It was extended linearly to cut regions
where the $\pi$C peak had insufficient yield.

The $\pi p$ scattering yield was extracted from the $E_{\mathrm{loss}}$ 
spectra for a wide range of cuts on the target signal size. The region 
of the $\pi p$ peak was fitted by a Gaussian peak and a flat distribution 
describing the remaining background under the peak. 
A systematic error of the analyzing
power due to the description of the 
$E_{\mathrm{loss}}$ distributions by the fit function was 
derived from the change of the analyzing power with the applied cut
on the active target signal, which is sensitive to an incorrect
description of the remaining background distribution (see next section).

\subsection{Analyzing power}

The analyzing power was calculated from the normalized $\pi p$ yields
(i.e. relative cross sections) 
for all runs in one set by using the linear relation between polarization
and cross section given in Eq. 1. The normalized yields were plotted against 
polarization and fitted by a straight line. The analyzing power was
then calculated from the line parameters.

As a consistency check for the calibration and yield extraction procedures,
the analyzing power was calculated for a wide range of cuts on the 
target signal size. Fig. 2 shows typical results. The top panel shows the 
dependence of the analyzing power on the required minimum target signal
size. There is no significant dependence at all cut levels.
The middle panel shows the corresponding reduced $\chi^2$ for the 
straight-line fit to the relative cross section. This quantity tests the 
internal consistency of the runs. Finally,
the bottom panel shows three selected fits of relative cross sections.
The corresponding cuts are 1000, 1200, and 1400 in the units of the 
upper panels, from top to bottom. Good consistency is found. 

The statistical error of the calculated analyzing power depends only
weakly on the applied cut. This situation is due to the fact
that the smaller number of counts for higher cuts is at least 
partly compensated by a lower background level. The final value of the
analyzing power was taken at the minimum of the statistical error.

\section{Results and discussion}

The results are listed in Tab. 1 and plotted in Figs. 3 and 4 
for $\pi^+ p$ and $\pi^- p$ scattering, respectively. 
Provided in the table are the values of the analyzing power 
with the statistical error as well as the systematic
error from the relative normalization, the active target ADC calibration,
and the fit to the $E_{\mathrm{loss}}$ spectra. The three sets are 
independently subject to an overall 3.5\% relative normalization error 
due to the systematic uncertainty in the polarization measurement.

The current results are compared with predictions of the KH80 \cite{KH80} 
and SAID \cite{SAID}
phase shift analyses and with previous data \cite{Wieser} in Fig. 3. 
The solid and dashed lines represent the FA02 \cite{SAID} and KH80
\cite{KH80} phase shift predictions, respectively. 
The data from Wieser et al. 
\cite{Wieser} at 68.3 MeV are shown as stars, the solid 
points represent the results from the current work (only statistical
errors are shown).

The phase shift results are in agreement with the data at 
87.2 and 77.2 MeV, with a somewhat better description by the
KH80 prediction at the lower energy. At 68.5 MeV, the predictions 
are significantly above the data in the forward-angle region, 
which is in agreement
with the findings from the earlier measurement \cite{Wieser}.
At 57.2 MeV, we again find agreement between the phase shift 
predictions and the data. The predictions are somewhat above the 
data at large angles at the two lowest energies. 
For the three measured points for $\pi^- p$ scattering in Fig. 4, 
the present
data are in agreement with previous data \cite{Patterson} 
and the phase-shift prediction results within the error bars 
(which are substantially larger than for the $\pi^+ p$ data). 

Overall, we do not
observe strong deviations of the phase shift predictions from 
the experimental results. 
This observation is similar to the findings for low energy
analyzing powers in $\pi^- p$ elastic scattering \cite{Patterson}.
Further experimental
information is expected from measurements of differential
cross sections in the Coulomb-nuclear interference region
down to very low energies (20 MeV)\cite{E778}, which have
been taken by the CHAOS collaboration and are currently 
being analyzed. If no significant deviations from the phase shift 
analyses are observed, then alternative explanations
for the large values of the sigma term from recent extractions
\cite{Marcello1} will have to be found. Several ideas have been 
put forward, questioning the
extraction procedure from $\pi p$ data \cite{Stahov}, the equivalence of 
the (sigma) terms extracted from $\pi p$ scattering and baryon masses
\cite{Gibbs2}, or the validity of the connection \cite{Gasser} of 
baryon masses and sigma term \cite{Thomas}.

\subsection{Acknowledgment}
This work was supported by the German minister of education
and research (BMBF) under contract 06TU201  and the Deutsche
Forschungsgemeinschaft (DFG: Europ. Graduiertenkolleg, 
Heisenberg-Programm).


\begin{table}
\begin{tabular}{|c|c|c|c|c|c|c|} \hline
Set & Energy (MeV)& Beam & $\Theta_{cm}$ (deg)& ~~~$A_y$~~~ & $\Delta A_y^{stat}$ & $\Delta A_y^{syst}$ \\
\hline
1   &   87.2     & $\pi^+$ &  48.2              & 0.378 & 0.054            & 0.031\\
   &       &  &  65.4              & 0.406 & 0.023             & 0.017\\
   &       &  &  81.8              & 0.392 & 0.015             & 0.012\\
   &       &  &  97.3              & 0.214 & 0.018             & 0.006\\
   &       &  & 112.1              & 0.153 & 0.014             & 0.011\\
   &       &  & 126.1              & 0.055 & 0.013             & 0.013\\
   &  68.4      &  &  71.1             & 0.317 & 0.036              & 0.016\\
   &       &  & 101.5              & 0.136 & 0.024             & 0.007\\
   &   57.2     &  &  58.6             & 0.344 & 0.045              & 0.015\\
   &       &  &  69.7             & 0.317 & 0.026              & 0.015\\
   &       &  &  85.7              & 0.264 & 0.028             & 0.025\\
   &       &  & 101.1             & 0.154 & 0.022              & 0.008\\
\hline
2   &   51.2     & $\pi^+$ &  63.9              & 0.340 & 0.053            & 0.023\\
   &       &  &  85.4             & 0.183 & 0.032             & 0.018\\
   &       &  &  100.8            & 0.088 & 0.018               & 0.012\\
   &       &  & 115.6            & 0.051 & 0.022               & 0.016\\
   &   45.2     &  &  85.2             & 0.144 & 0.026              & 0.012\\
   &       &  & 100.6              & 0.053 & 0.027             & 0.009\\
\hline
3   &   87.2     & $\pi^+$ &  66.5              & 0.411 & 0.019            & 0.007\\  
   &       &  &  88.1             & 0.339 & 0.018             & 0.010\\
   &   77.2     &  &  63.0              & 0.456 & 0.021             & 0.014\\
   &       &  &  84.7              & 0.353 & 0.019             & 0.008\\
   &   68.6     &  &  59.5             & 0.375 & 0.023              & 0.013\\
   &       &  &  81.3             & 0.317 & 0.018              & 0.010\\
   &   57.3     &  &  75.8             & 0.302 & 0.024              & 0.014\\
   &   87.2     & $\pi^-$ &  60.9              & 0.128 & 0.048             & 0.019\\
   &       &  &  90.7              & 0.088 & 0.050             & 0.028\\
   &   67.3     &  &  83.5           & 0.092 & 0.048                & 0.016\\
\hline
\end{tabular}

\caption{Results of this measurement. The systematic errors 
include errors from the relative normalization, the active target 
ADC calibration, and the fit to the $E_{\mathrm{loss}}$ spectra. The three 
sets are independently subject 
to an overall 3.5\% relative normalization error due to the systematic
uncertainty in the polarization measurement.
}

\label{restable}
\end{table}

\begin{figure}
\begin{center}
\includegraphics[width=8cm]{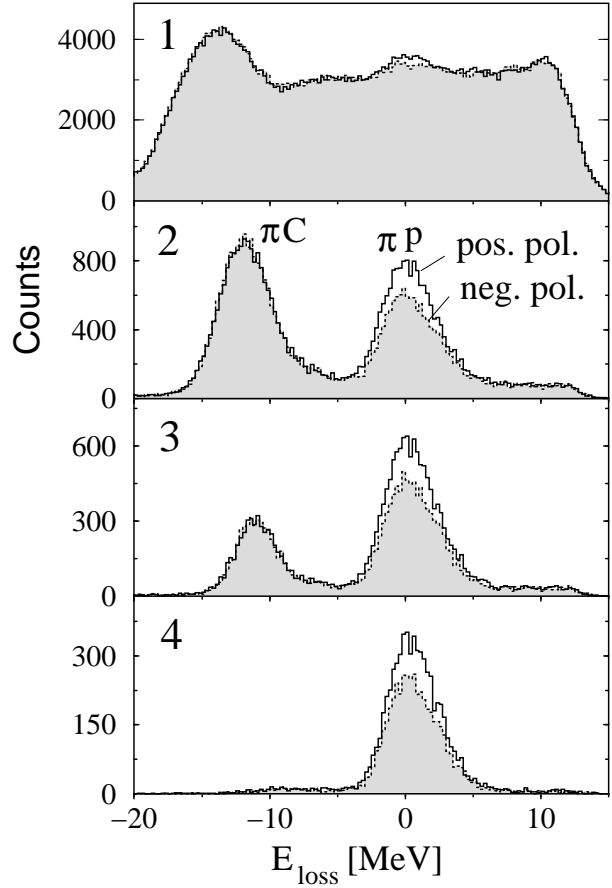}
\caption{LEPS focal plane $\pi^+$ spectra at $T_\pi$ = 68.6 MeV,
$\Theta_{cm}$ = 81.3 deg. for positive (solid line) and 
negative (dashed line, shaded area) target polarization, 
for different conditions 1 to 4. Plot 1 shows spectra with 
spectrometer information only, and no required active target
signal. Plots 2 to 4 show the changes of the spectra as 
increasing minimum signal sizes of the active target are
imposed. The expected positions of $\pi$ carbon and 
$\pi$ proton elastic scattering are labelled in panel 2.}
\end{center}
\end{figure}

\begin{figure}
\begin{center}
\includegraphics[width=8cm]{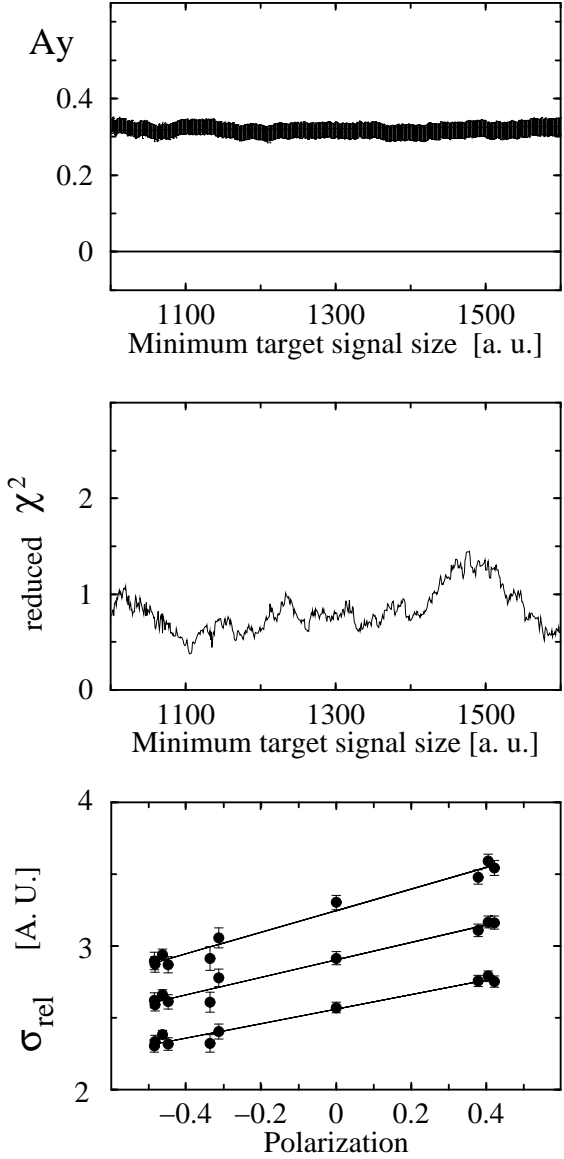}
\caption{Analyzing power extraction at $T_{\pi^+}$ = 68.6 MeV,
$\Theta_{cm}$ = 81.3 deg. The top panel shows  the 
dependence of the analyzing power on  the required minimum target signal
size. The result is consistent with a constant for all cut levels.
The middle panel shows the corresponding reduced $\chi^2$ for the straight
line fit to the relative cross section. 
The bottom panel shows three selected fits of relative cross sections
over polarization. The corresponding cuts are 1000, 1200, and 1400 in 
the units used in the upper panels, from top to bottom.}
\end{center}
\end{figure}

\begin{figure}
\begin{center}
\includegraphics[width=8cm]{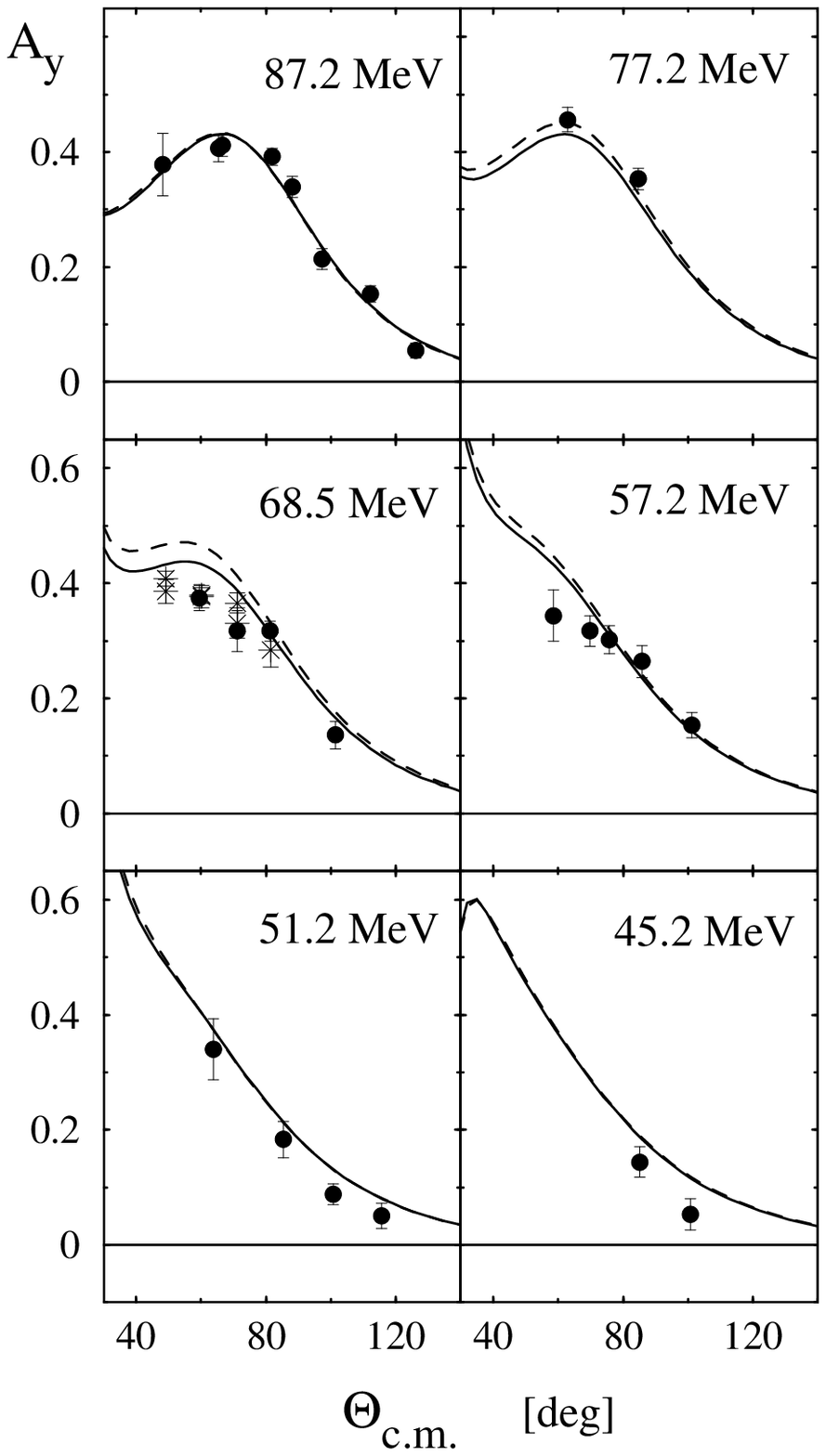}
\caption{Angular distributions of analyzing powers for
$\pi^+ p$ elastic scattering. The lines
represent phase shift predictions from KH80 \cite{KH80} (dashed)
and FA02 \cite{SAID} (solid). Previous data \cite{Wieser} are shown as
stars, results from this experiment as solid points.}
\end{center}
\end{figure}

\begin{figure}
\begin{center}
\includegraphics[width=8cm]{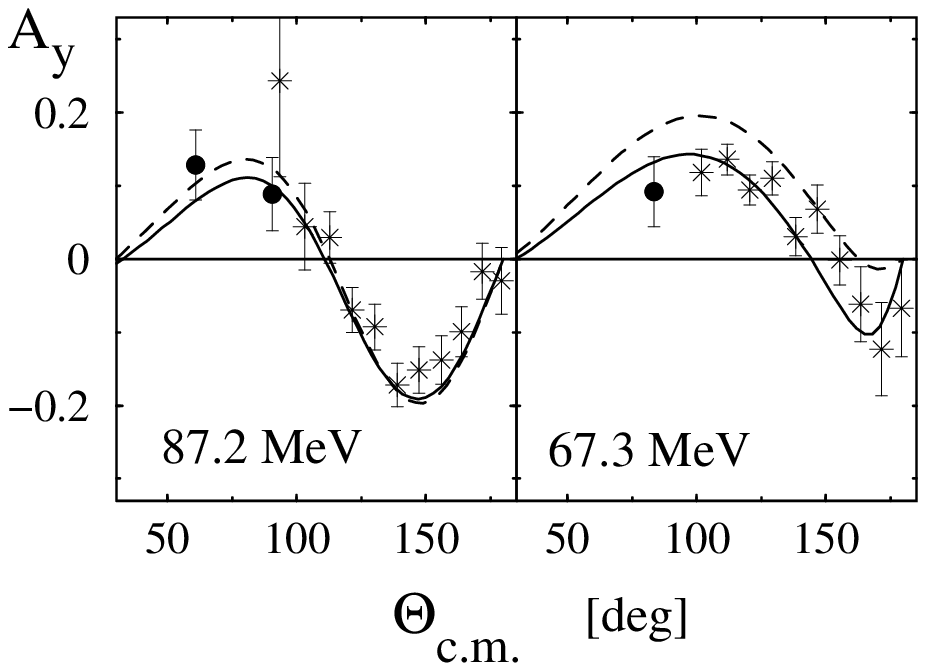}
\caption{Angular distributions of analyzing powers for
$\pi^- p$ elastic scattering. The lines
represent phase shift predictions from KH80 \cite{KH80} (dashed)
and FA02 \cite{SAID} (solid). Previous data \cite{Patterson} are shown as
stars, results from this experiment as solid points.}
\end{center}
\end{figure}


\begin{thebibliography}{99}
\bibitem{chi2000} Ulf-G. Mei\ss ner and G.R. Smith in 
{\it Proceedings from the Institute for Nuclear Theory-Vol.11}, 329, Editors
A.M. Bernstein, J.L. Goity, U.-G. Mei\ss ner, World Scientific (2001).
\bibitem{Koch} R. Koch, Z. Phys. {\bf C15}, 161 (1982).
\bibitem{KH80} R. Koch and E. Pietarinen, Nucl. Phys. 
{\bf A336}, 331 (1980); G. H\"ohler, in {\it Pion Nucleon Scattering}, edited
by H. Schopper, Landolt-B\"ornstein, New Series, Group X, Vol. I, 9b2
(Springer Verlag, Berlin, 1983). 
\bibitem{Gasser} J. Gasser, Ann. Phys. (N.Y.) {\bf 136}, 62 (1981).
\bibitem{Borasoy} B. Borasoy and Ulf-G. Mei\ss ner, Ann. Phys. (N.Y.) {\bf 254}, 
192 (1997).
\bibitem{Marcello1} M.M. Pavan, R.A. Arndt, I.I Strakovsky, and R.L. Workman, 
in {\it Proceedings of the 9th International
Symposium on Meson-Nucleon Physics and the Structure of the Nucleon,
Washington D.C.}, edited by H. Haberzettl and W.J Briscoe [$\pi$N
Newsletter {\bf 16}, 110 (2002)]. 
\bibitem{Gutsche} V.E. Lyubovitskij, T. Gutsche, A. Faessler and E.G. Drukarev,
Phys. Rev. D {\bf 63}, 054026 (2001).
\bibitem{Bazarko} A.O. Bazarko et al., Z. Phys. {\bf C65}, 189 (1995).
\bibitem{Gibbs} W.R. Gibbs, Li Ai, and W.B.Kaufmann, Phys. Rev. Lett. {\bf 74}, 3740 (1995).
\bibitem{Matsinos} E. Matsinos, Phys. Rev. C {\bf 56},3014 (1997).
\bibitem{fettes} N. Fettes and Ulf-G. Meissner, Phys. Rev. C {\bf 63}, 045201 (2001).
\bibitem{Marcello} M.M. Pavan et al., Phys. Rev. C {\bf 64}, 064611 (2001).
\bibitem{Sevior} M.E. Sevior et al., Phys. Rev. C {\bf 40}, 2780 (1989).
\bibitem{Hofman1} G.J. Hofman et al., Phys. Rev. C {\bf 58}, 3484 (1998).
\bibitem{Hofman2} G.J. Hofman et al., Phys. Rev. C {\bf 68}, 018202 (2003).
\bibitem{FettesMatsinos} N. Fettes and E. Matsinos, Phys. Rev. C {\bf 55},
464 (1997).
\bibitem{Wieser} R. Wieser et al., Phys. Rev. C {\bf 54}, 1930 (1996).
\bibitem{Patterson} J.D. Patterson et al., Phys. Rev. C {\bf 66}, 025207 (2002).
\bibitem{KF} K. F\"ohl, PhD thesis, University of T\"ubingen (1996).
\bibitem{Ben} B. van den Brandt et al., Nucl. Instr. Meth. A {\bf 446},
  592 (2000).
\bibitem{SAID} R.A. Arndt, I.I. Strakovsky, R.L. Workman, M.M. Pavan, Phys. Rev. C {\bf 52}, 2120 (1995);
               R.A. Arndt, W.J. Briscoe, I.I. Strakovsky, R.L. Workman, M.M. Pavan, arXiv:nucl-th/0311089v2 (2004);
               the SAID analysis is available at http://gwdac.phys.gwu.edu/
\bibitem{E778} H. Denz, in {\it Proceedings of the 9th International
Symposium on Meson-Nucleon Physics and the Structure of the Nucleon,
Washington D.C.}, edited by H. Haberzettl and W.J Briscoe [$\pi$N
Newsletter {\bf 16}, 302 (2002)]. 
\bibitem{Stahov}
J. Stahov, arXiv:hep-ph/0206041 (2002).
\bibitem{Gibbs2} W.R. Gibbs, Mod. Phys. Lett. {\bf A18}, 1171 (2003). 
\bibitem{Thomas} D.B. Leinweber, A.W. Thomas and S.V. Wright,
Phys. Lett. {\bf B482}, 109 (2000).

\end{thebibliography}
\end{document}